# SPIN TRACKING AT THE ILC POSITRON SOURCE[*]

V. KOVALENKO[†1], O.S. ADEYEMI[1], A. HARTIN[3], G. A. MOORTGAT-PICK[1,3],
L. MALYSHEVA[1], S. RIEMANN[2], F. STAUFENBIEL[2], A. USHAKOV[1].

[1]*II. Institute for Theoretical Physics, University of Hamburg,*
*Hamburg, 22607, Germany*

[2]*II. Deutsches Electronen-Synchrotron, DESY*
*Zeuthen, 15738, Germany*

[3]*II. Deutsches Electronen-Synchrotron, DESY*
*Hamburg, 22607, Germany*

In order to achieve the physics goals of future Linear Colliders, it is important that electron and positron beams are polarized. The baseline design at the International Linear Collider (ILC) foresees an e+ source based on helical undulator. Such a source provides high luminosity and polarizations. The positron source planned for ILC is based on a helical undulator system and can deliver a positron polarization of 60%. To ensure that no significant polarization is lost during the transport of the e- and e+ beams from the source to the interaction region, precise spin tracking has to be included in all transport elements which can contribute to a loss of polarization, i.e. the initial accelerating structures, the damping rings, the spin rotators, the main linac and the beam delivery system. In particular, the dynamics of the polarized positron beam is required to be investigated. In the talk recent results of positron spin tracking simulation at the source are presented. The positron yield and polarization are also discussed depending on the geometry of source elements.

## 1. Introduction

The undulator scheme of polarized positron production was proposed by Michailichenko and Balakin in 1979 [1] and has been chosen as a baseline for the International Linear Collider (ILC) (see Figure 1). The scheme is based on a two stage process, where at the first stage the circularly polarized photons are generated in a helical magnetic field and then, at the second stage, these photons are converted into longitudinally polarized positrons and electrons in a thin target. The circular polarization of the photons is transferred to longitudinal polarization of the electrons and positrons. The main parts of the positron source are: the helical undulator, the photon collimator, the target, the optical matching device installed after the target in order to capture the longitudinally polarized positrons, the RF section embedded in a solenoid to capture and pre-accelerate the beam up to 125 MeV, then the pre-acceleration to 400 MeV, and after that the booster linac accelerates the beam to 5 GeV. In order to preserve the polarization of the beam in the damping ring (DR) the spin orientation of the positrons has to be rotated from the longitudinal into the vertical direction before the damping ring via a spin rotator.

---

[*] This work is supported by the German Federal Ministry of Education and Research, Joint Research Project R&D Accelerator "Spin Management", contract number 05H10GUE.
  Talk was presented at POSIPOL 2011 conference.
[†] e-mail: valentyn.kovalenko@desy.de

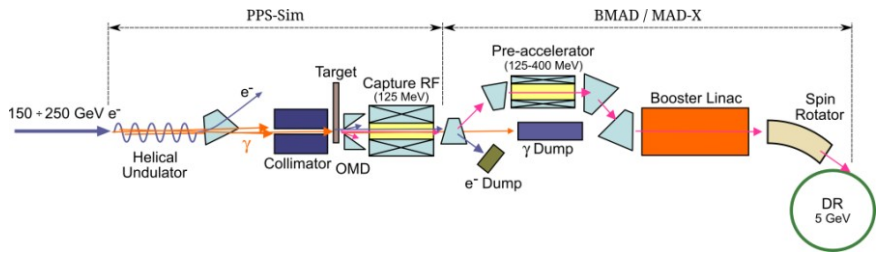

Figure 1. Schematical layout of polarized positron source based on undulator scheme.

The efficiency of the beam generation as well as the beam optics downstream the target plays the crucial role for the design of the positron source components. In this study we investigate a quarter-wave transformer (QWT) as an optical matching device in order to find the optimal geometry parameters to fulfill the beam yield requirements also for high degrees of positron beam polarization.

## 2. Polarized Positron Source Simulation (PPS-Sim) Code

There are several Monte-Carlo tools, for example FLUKA and EGS which are used to perform positron production simulations. But none of them allows to calculate particle beam dynamics in the accelerating structure, because the electrical field is not implemented. At the same time there are a numerous well developed codes for simulations of beam dynamics in the accelerating structures of linear and circular accelerators (MAD-X, PARMELA, BMAD, Elegant, etc.). However all these codes do not include the positron production process and require an input file from FLUKA or EGS and some of them do not take into account spin of particles.

Therefore, we used the program code PPS-Sim [2, 3] which is based on Geant4 including positron production, energy deposition and also the transport of charged particles in magnetic and electric fields, and the spin transport. PPS-Sim is an ideal tool to combine beam generation, beam focusing and particle acceleration taking into account the spin of the particles.

## 3. Positron Source Parameters and Simulation Results

The positron source parameters for the ILC used for the simulations are presented in Table 1.

Table 1. ILC Positron source parameters.

| Electron beam energy | < 250 GeV |
|---|---|
| Number of positrons | $3 \cdot 10^{10}$ e$^+$/bunch |
| Number of bunches | 2625 or 1312 bunches/train |
| Repetition rate | 5 Hz |
| Undulator K-value | 0.92 |
| Undulator period | 11.5 mm |
| Undulator length | 231 m |
| Undulator-Target Distance | ~ 500 m |
| Target material | Ti6Al4V |
| Target thickness | 0.4 $X_0$ |
| Target rotation speed | 100 m/s |
| OMD | QWT |
| DR acceptance: energy spread | 1 % |
| DR acceptance: emittance, $\varepsilon_{nx}+\varepsilon_{ny}$ | 0.09 rad m |
| DR acceptance: long. bunch size | 34.6 mm |

The circularly polarized photons hit Ti alloy target with a thickness of 0.4 radiation length and produce longitudinally polarized positrons. Then the generated positron beam is collected and accelerated to 125 MeV. The optical matching device is a QWT and consists of three solenoids (see Figures 2). The first two solenoids (bulking and focusing) have a higher magnetic field than the third one that is called background solenoid. The OMD is followed by a 1.3 GHz accelerating cavity embedded into solenoid with constant B-field. The E-field of the RF cavity is modeled as harmonic function. PPS-Sim does not include the whole beam line up to the DR at 5 GeV. In order to estimate the number of positrons out off the DR acceptance, PPS-Sim applies cuts on the longitudinal bunch size and on the sum of x- and y-emittances. The model of the positron source with quarter-wave capturing is shown in Figure 3.

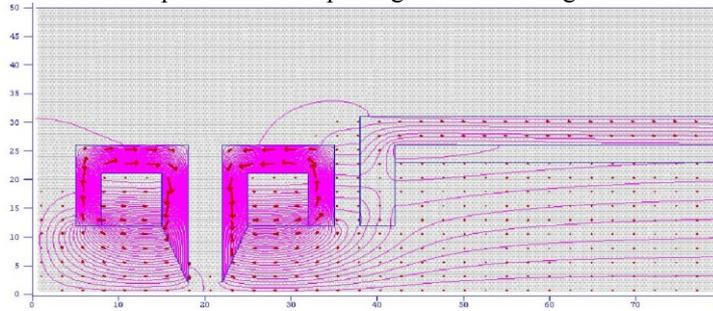

Figure 2. Magnetic field distribution in the quarter-wave transformer.

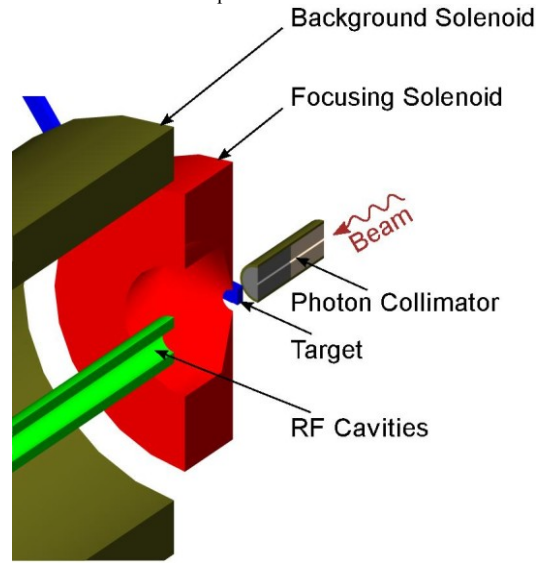

Figure 3. Model of quarter-wave transformer capturing for positron source. Only the focusing magnet and background solenoid of the QWT are shown.

### 3.1. *Results and discussion*

In Figure 4 it is shown how the polarization and yield of positrons depend on the magnetic field of QWT for different drive beam energies. In the case of a 250 GeV drive beam the polarization lies in a range of 24-26%. Different magnetic fields of the focusing solenoids of QWT (1÷2T) do not significantly affect the polarization. This value of polarization degree is corresponding to the current baseline design not sufficient to achieve the full physics potential with polarized beams. The positron yield is required to be 1.5 $e^+/e^-$. It should be noted that the yield is calculated as the ratio of captured positrons to electrons in the drive beam.

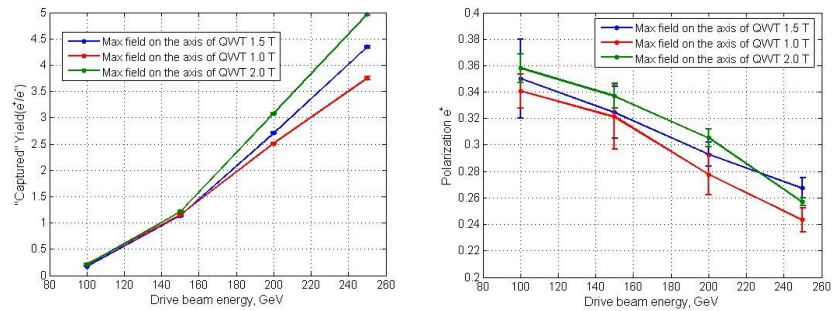

Figure 4. Polarization and yield of positrons versus different magnitudes of the QWT magnetic field for different drive beam energies.

One method to increase the positron polarization degree is to apply a photon collimator. Let us consider 250 GeV beam and how the photon collimator with different radii affects the positron yield and polarization (see Figure 5). Obviously smaller radius of the collimator results in higher polarization and lower yield. The photon collimator with 1 mm radius aperture increases the positron polarization up to approximately 60% which is agreed to be the goal. However, one has to be deal whether heat loading, energy deposition or other factors might lead to destruction of inner part of the collimator. Hence, the results presented below will consider the case with 2 mm collimator radius. For a 250 GeV electron beam a polarization of about 31-34% can then be achieved.

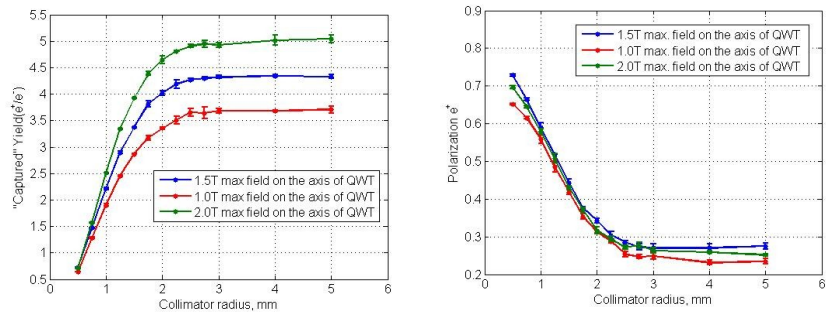

Figure 5. Polarization and yield of positrons versus different radii of the photon collimator for a drive beam energy of 250 GeV.

Figure 6 shows the polarization and yield depending on the distance between the QWT and the target. We study the initial conditions:
- Drive beam energy 250 GeV
- K=0.92, λ=11.5 mm
- No collimation
- Distance between undulator center and QWT ~500 m
- Undulator length is 231 m
- Length of QWT is 130 mm
- Maximal magnetic field on the axis of QWT is 1 T

It should be noted that the phase of the RF field is optimized to get a higher value of positron yield. If we place the QWT at 10 mm from the target we will get a maximum yield and a minimum polarization of about 23%. Increasing the distance we lose particles and the yield goes down but at the same time polarization grows up. For example, if we place the QWT at 150 mm from the target it is possible to get 28% positron polarization still providing the yield of 3 positrons per electrons.

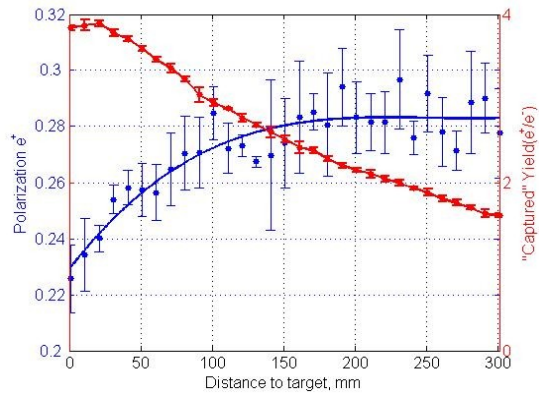

Figure 6. Polarization and yield of positrons versus distance between QWT and target without collimation for 250 GeV drive beam energy.

Using the same conditions mentioned above and applying a photon collimator with a radius of 2 mm in addition, a polarization enhancement of 6-7% can be observed. At the same time the yield still fulfils the requirement of 1.5 positrons per electrons (see Figure 7).

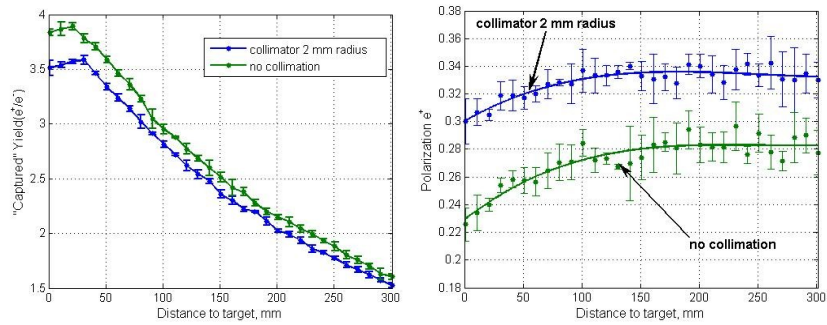

Figure 7. Polarization and yield of positrons versus distance between QWT and target with 2 mm radius of photon collimation for 250 GeV drive beam energy.

We also changed the length of the QWT in our simulations. In Figure 8 the corresponding dependencies are presented. The photon collimator was not applied in this case. It can be observed that the polarization only increased up to 24.5% caused by lengthening of QWT. The optimum yield is achieved for QWT lengths of 110 – 120 mm. The polarization in this case is 23%.

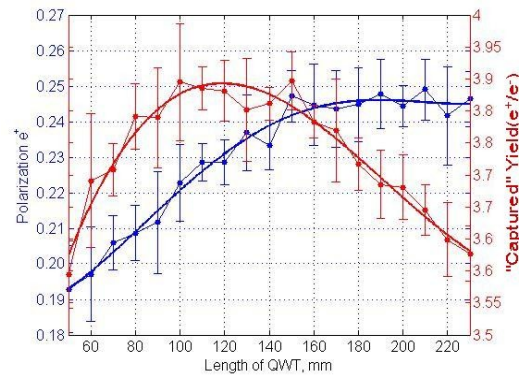

Figure 8. Polarization and yield of positrons versus length of QWT without collimator for 250 GeV drive beam energy.

## 4. Summary

For a 250 GeV electron beam and without photon beam collimation, the positron polarization lies in a range of 24-26%. This value of polarization might not be sufficient to achieve the full physics potential of the ILC with polarized beams. Increasing the distance between the target and the QWT gives an enhancement of polarization by 6-7%. A photon collimator with 2 mm aperture radius increases the polarization up to 35% where the positron yield still fulfils the requirement of 1.5 e+/e-. A longer QWT also slightly increases the value of polarization in addition.